\documentclass[12pt]{iopart}
\usepackage{graphicx}
\usepackage{amssymb}
\usepackage{iopams}
\usepackage{setstack}

\begin{document}
\title{Sub shot-noise interferometry from measurements of the one-body density}
\author{J Chwede\'nczuk$^1$, P Hyllus$^2$, F Piazza$^3$ and A Smerzi$^{4,5}$}
\address{
  $^1$Faculty of Physics, University of Warsaw, ul. Ho\.za 69, 00-681 Warsaw, Poland\\
  $^2$Department of Theoretical Physics, University of the Basque Country UPV/EHU,
  P.O. Box 644, E-48080 Bilbao, Spain\\
  $^3$Physik Department, Technische Universit\"at M\"unchen, 85747 Garching, Germany\\
  $^4$INO-CNR  and LENS, 50125 Firenze, Italy \\
  $^5$INO-CNR BEC Center and Dipartimento di Fisica, Universit\'a di Trento, 38123 Povo, Italy
}
\begin{abstract}
  We derive the asymptotic maximum-likelihood phase estimation uncertainty for any interferometric protocol where the positions of the probe particles 
  are measured to infer the phase, but where correlations between the particles are not accessible.
  First, we apply our formula to the estimation of the phase acquired in the Mach-Zehnder interferometer and recover the well-know momentum formula for the phase sensitivity.
  Then, we apply our results to interferometers with two spatially separated modes, 
  which could be implemented with a Bose-Einstein condensate trapped in a double-well potential. 
  We show that in a simple protocol which estimates the 
  phase from an interference pattern a sub shot-noise phase uncertainty of up to $\Delta\theta\propto N^{-2/3}$ can be achieved. 
  One important property of this estimation protocol is that its sensitivity does not depend on the value of the phase $\theta$, contrary to the sensitivity given by the momentum
  formula for the Mach-Zehnder transformation.
  Finally, we study the experimental implementation of the above protocol in detail, by numerically simulating the full statistics 
  as well as by considering the main sources of detection noise, and argue that the shot-noise limit could be surpassed with current technology.
\end{abstract}
\maketitle

\section{Introduction}

Quantum interferometry, a rapidly developing branch of modern physics,
exploits some striking features of quantum mechanics in order to build ultraprecise measuring devices \cite{giovanetti}. 
It employs non-classical states of light \cite{light,light1,light2,light3} or matter \cite{cold1,cold2,cold3,gross,esteve,riedel, maussang} 
to beat the shot-noise limit (SNL) $\Delta\theta=1/\sqrt{N}$ -- the limit of 
precision of parameter estimation of a phase shift $\theta$ set by classical physics.
Here $N$ is the number of particles in the probe state.
Recent achievements in the preparation of entangled states of atoms have put atomic interferometry \cite{cronin}
in the \emph{avant-garde} for precise determination of electromagnetic 
\cite{ketterle_casimir,vuletic_casimir,cornell} and gravitational \cite{kasevich,fattori,hinds} interactions. 

A paradigmatic example and benchmark for every interferometer is the Mach-Zehnder interferometer 
(MZI), where two ``beams'' propagating along separated paths accumulate a relative phase $\theta$ 
(to be estimated) and are subsequently recombined through a beam splitter. 
A similar protocol which uses internal states instead of separate
paths is known as Ramsey-spectroscopy \cite{wine}. It
has been recently realized with a Bose-Einstein condensate (BEC), 
employing two hyperfine states of the atoms \cite{gross}. 
The beam-splitter was mimicked by coupling the two
modes with  a micro-wave pulse for a precisely chosen amount of time.
Atomic interferometers with spatially separated modes, which could for instance be used 
to measure forces decaying with the distance
are more challenging in implementation due to the difficulty of performing
the beam-splitter transformation \cite{pezze_BS, grond_MZI}.

In this paper we consider a simple ``double-slit'' interferometer in which two beams are recombined through a free expansion of two initially localized clouds with
the output signal obtained by measuring the positions of the particles.
After preparation of a suitably entangled input state, sub shot-noise (SSN) interferometry requires thus refined particle detection at the output.
Some new techniques of atom-position measurements, like the
micro-channel plate \cite{jeltes}, the tapered fiber \cite{heine}, the light-sheet method \cite{perrin_sheet} or techniques involving atomic fluorescence from the lattice \cite{sherson}, 
give hope for an almost 100\% efficient single-atom detection in the nearby future. 
Such a tool in principle could even give access to atom-atom correlations at all orders. 

It has been shown \cite{chwed,piazza} that the measurement 
of the $N$-th order correlation function
is the best possible estimation strategy for inferring the phase between two interfering BEC wave-packets, and allows to reach the Heisenberg limit of phase uncertainty. 
However, even with small BECs, the measurement of the $N$-th order correlation function would require substantial experimental effort, 
since a huge configurational space must be probed with sufficient signal to noise ratio.

\begin{figure}[htb]
  \includegraphics[scale=0.4,clip]{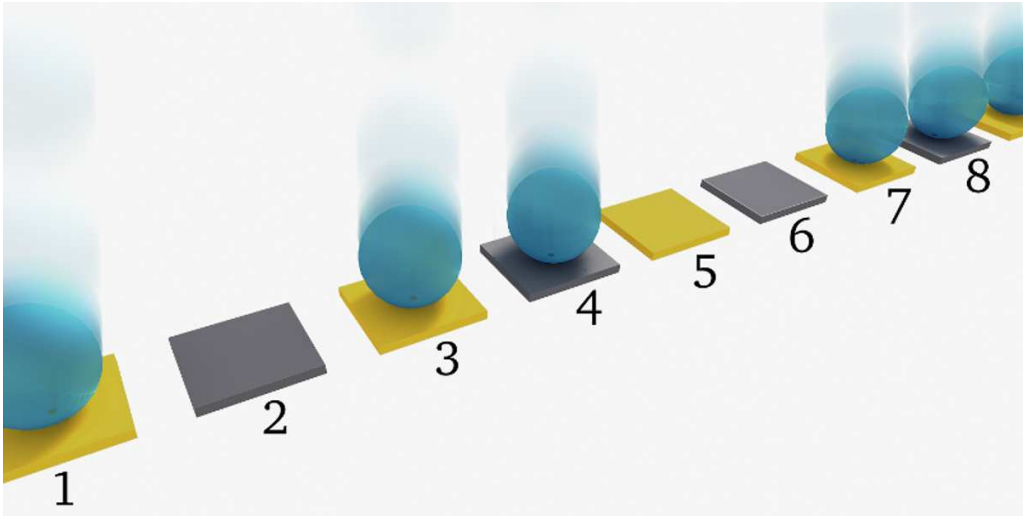}
  \caption
      {Schematic representation of the atom position measurement, where the detectors (boxes) turn yellow (no. 1, 3  and 7) when they are hit by atoms (spheres). 
        Some detectors however remain gray (as the detector no. 8), so the detection efficiency drops, 
        and sometimes a neighbouring box turns yellow (no. 5 instead of no. 4), which limits the spatial resolution.}
      \label{scheme}
\end{figure}

The difficulty of measuring high order correlation functions is the motivation for this work. 
We show that by measuring positions of particles the phase can be estimated with a SSN phase uncertainty using the simple one-body density.
We consider two possible detection scenarios: the output signal consists of (a) the positions of single atoms, or (b) the number of atoms per pixel, which corresponds to 
the commonly employed least-squares estimation from the fit to the average density
\cite{chwed2}.
For both cases, we compute the general asymptotic phase uncertainty of a 
maximum-likelihood phase estimation scheme using the one-body density only.
 
We verify that in an MZI situation, estimating the phase from the
one-body density is equivalent to estimation from the average population imbalance
between the arms of the interferometer, 
and recover the known result that the sensitivity can saturate the Heisenberg limit.

We then apply this estimation protocol to our case of interest, namely,
two interfering BEC wavepackets.
We provide an analytical expression for the phase uncertainty which shows SSN scaling with phase squeezed states in input. 
Contrary to the MZI, the phase uncertainty does not depend on the phase $\theta$, which could be an important advantage.
We then analyze the full statistics of the phase estimation by numerically simulating experiments with a realistic number of particles. 
We find that already small statistical samples are sufficient to saturate the analytical asymptotic prediction for the phase uncertainty.
Finally, we discuss the main sources of noise affecting the interferometric precision, which we expect to concern the atom detection stage.
We argue that including the effect of imperfect detection the SNL can still be surpassed,
and that the necessary amount of squeezing could be achieved in a realistic 
double-well setup. 

The paper is organized as follows. In Section \ref{est} we derive the general expression for the uncertainty of the phase estimated from the one-body density.
We consider two possible detection scenarios. We assume either having access to positions of single particles or
a coarse grained measurement due to a limited detection capability. 
In Section \ref{MZI}, we apply the above results to the case of the MZI. 
In Section \ref{interf} we turn to our case study where the interferometer consists of a simple phase-imprint between the modes followed by the 
ballisitic expansion of the mode functions. 
We also show how the phase uncertainty for the phase estimation from the density of the inteference pattern is influenced by characteristic sources of noise. 
Some details of analytical calculations are presented in the Appendices.

\section{Sensitivity of the one-body density estimator}
\label{est}

An interferometric estimation protocol can be divided in: (i) the interferometer transformation, 
(ii) the measurement at the output, and (iii) the phase inference through an estimator.
In (i), the interferometer imprints a relative phase $\theta$ on the $N$-particle input state $|\psi_{\mathrm{in}}\rangle$. Such transformation
can be represented by a unitary evolution operator $\hat U(\theta)=e^{i\theta\hat h}$, where $\hat h$ is linear 
-- i.e. can be written as a sum of operators acting on each particle separately -- and does not depend on $\theta$.
In the Heisenberg picture, the field operator evolves as $\hat\Psi(x|\theta)\equiv\hat U^\dagger(\theta)\hat\Psi(x)\hat U(\theta)$ and $|\psi_{\mathrm{in}}\rangle$ is unchanged.
In the following, we will specify our arguments to the case of particle position measurements at the output, although the results of Sec.~\ref{sing_at_det} are valid in general.
We will thus consider next that in stage (ii), upon leaving the interferometer, the positions of the particles are detected. 
As observed above, due to the difficulty of obtaining high order correlation functions, 
often only the lowest, namely the density, can be precisely measured. 
This density, when normalized, gives 
\begin{equation}\label{single}
  p_1(x|\theta)=\frac{\big\langle\hat\Psi^\dagger(x|\theta)\hat\Psi(x|\theta)\big\rangle}N,
\end{equation}
the probability density of measuring a single particle at position $x$ given $\theta$.
The average value is calculated for the input state $|\psi_{\mathrm{in}}\rangle$.
Next, we assume having no access to the correlations, so the phase is inferred only using $p_1(x|\theta)$ and the measurement outcomes obtained in stage (ii). 

\subsection{Single-atom detection}
\label{sing_at_det}

When in (ii) single-atom detection is performed, see Fig. \ref{scheme}, 
each experiment gives the positions of the $N$ atoms, $\vec x_N^{(i)}=x_1^{(i)},\ldots, x_N^{(i)}$, 
where the label $i=1,\dots,m$ indicates the particular iteration of the experiment. The accumulated set of data, $N$ times $m$ positions, is used to construct the likelihood function,
\begin{equation}
  \mathcal{L}(\varphi)=\prod_{i=1}^m\prod_{k=1}^Np_1(x^{(i)}_k|\varphi). \label{like}
\end{equation}
The prescription of the Maximum Likelihood Estimator (MLE) is to infer the phase $\theta^{(m)}_{\mathrm{ML}}$ in stage (iii) as the value of $\varphi$ which maximizes Eq.~(\ref{like}). 
As demonstrated in the \ref{ap_lik}, this MLE is consistent, i.e $\theta^{(m)}_{\mathrm{ML}}\rightarrow\theta$ for $m\rightarrow\infty$.
There, we also obtain the uncertainty of the estimator equal to
\begin{equation}
  \Delta^2\theta^{(m)}_{\mathrm{ML}}=\frac1m\frac1{NF_1}\left(1+(N-1)\frac C{F_1}\right),\label{sens}
\end{equation}
where $C$ depends on the two-body probability density of detecting one particle at $x_1$ and 
the other at $x_2$, 
\begin{equation}
  p_2(x_1,x_2|\theta)=\frac{\big\langle\hat\Psi^\dagger(x_1|\theta)\hat\Psi^\dagger(x_2|\theta)\hat\Psi(x_2|\theta)\hat\Psi(x_1|\theta)\big\rangle}{N(N-1)},\label{double}
\end{equation}
and reads \cite{nota_der}
\begin{equation} \label{c}
  C=\int dx_1 dx_2\,p_2(x_1,x_2|\theta)\frac{\partial_\theta p_1(x_1|\theta)}{p_1(x_1|\theta)}\frac{\partial_\theta p_1(x_2|\theta)}{p_1(x_2|\theta)}.
\end{equation}
Furthermore $F_1$ is the Fisher information calculated with the probability $p_1$,
\begin{equation}
  F_1=\int\!\! dx\,\frac{\left[\partial_\theta p_1(x|\theta)\right]^2}{p_1(x|\theta)}.\label{f1}
\end{equation}
Equation (\ref{sens}) is the first important result of this paper. The phase, which is estimated from the single-body probability, depends on both $p_1$ and $p_2$.
If $C=0$ or neglected as in \cite{chwed}, Eq.~(\ref{sens}) provides a shot-noise limited phase uncertainty, since $F_1\leq1$ \cite{chwed}. Yet, $C$ can assume negative values, 
allowing for SSN phase uncertainty, as will be demonstrated below. 

We also underline the generality of the above result, which 
is valid for any quantum state, where the parameter $\theta$ is estimated from the one-body density. In analogy, if two-body correlations can be measured in an experiment,
four-body correlations would enter in the corresponding expression for the
asymptotic phase uncertainty.

\subsection{Multiple-atom detection}
\label{mult_at_det}

It is also important to consider the possibility that the detectors cannot resolve positions of individual particles.
In this case, we need to use the coarse-grained density and assume that in stage (ii) in the $i$-th experiment the number of atoms $n_k^{(i)}$ in each of the
$k=1\ldots n_{\rm bin}$ bins is  measured  (the bin size $\Delta x$ must be small to precisely sample the density variations). 
The measurement is repeated $m$ times and the phase is estimated from a
least square fit of the one-body density to the accumulated data. As discussed in detail in \cite{chwed2}, 
the phase uncertainty of such fit is equivalent to the phase uncertainty of the MLE with the likelihood function
$\mathcal{L}_{\rm fit}(\varphi)=\prod_{k=1}^{n_{\rm bin}}p(\bar n_k|\varphi)$, where $p(\bar n_k|\varphi)$ is the probability for detecting
$\bar n_k=\frac1m\sum_{i=1}^mn_k^{(i)}$ atoms on average
in the $k$-th bin. For large $m$, according to the Central Limit Theorem, the probability $p(\bar n_k|\varphi)$ is a Gaussian with a mean
$\langle n_k \rangle=\Delta x N p_1(x_k|\varphi)$ and Poissonian fluctuations, $\Delta^2 n_k=\langle n_k \rangle$ \cite{chwed}. 
For the phase uncertainty of this MLE, we obtain similarly as in Section \ref{sing_at_det} that
\begin{equation}
  \Delta^2\theta^{(m)}_{\rm fit}=\frac{\sum_{k=1}^{n_{\rm bin}} \frac{\left[\partial_\theta\langle n_k\rangle\right]^2}{\langle n_k \rangle}+
    \sum_{k\neq l=1}^{n_{\rm bin}}\sigma_{k,l}^2\frac{\partial_\theta\langle n_k\rangle}{\langle n_k \rangle}\frac{\partial_\theta\langle n_l\rangle}{\langle n_l \rangle}}
    {m\left[\sum_{k=1}^{n_{\rm bin}} \frac{\left[\partial_\theta\langle n_k\rangle\right]^2}{\langle n_k \rangle}\right]^2},\label{sens_fit}
\end{equation}
where 
\begin{equation*}
  \sigma_{k,l}^2=(\Delta x)^2N\left[(N-1)p_2(x_k,x_l|\theta)-N p_1(x_k|\theta) p_1(x_l|\theta)\right].
\end{equation*}
In the continuous limit $\Delta x\rightarrow0$, the formulas (\ref{sens}) and (\ref{sens_fit}) coincide. 



\section{Estimation from the one-body density with the MZI}
\label{MZI}

As a benchmark for the phase estimation protocol introduced in Sec.~\ref{est}, we first consider the case where in stage (i) the system acquires the phase $\theta$ in a MZI. The interferometric sequence of the MZI consists of three steps. First,
the initial two-mode state $|\psi_{\rm in}\rangle$ passes a beam-splitter. Then, a relative phase $\theta$ is imprinted between the modes. In the final stage of the interferometer,
another beam-splitter acts on the state.

It can be easily shown that the evolution operator of the whole MZI sequence is $\hat U(\theta)=e^{-i\theta\hat J_y}$ \cite{nota_J}. 
It is convenient to switch to the Heisenberg picture. The initial field operator reads 
$\hat\Psi(x)=\psi_a(x)\hat a+\psi_b(x)\hat b$, where $\hat a$/$\hat b$ annihilates a particle from mode $a/b$ and the corresponding spatial mode function
is $\psi_{a/b}(x)$. 
When passing the MZI, this operator is transformed as follows
\begin{equation}
  \fl\hat\Psi(\theta)=e^{i\theta\hat J_y}\hat\Psi(x)e^{-i\theta\hat J_y}=
  \psi_a(x)\left[\hat a\cos\frac\theta2-\hat b\sin\frac\theta 2\right]+\psi_b(x)\left[\hat b\cos\frac\theta2+\hat a\sin\frac\theta 2\right].\nonumber
\end{equation}
We use this result to calculate the one-body [cf. Eq.~(\ref{single})] and two-body 
[cf. Eq.~(\ref{double})] probability densities that enter Eq.~(\ref{sens}), assuming that both the mode functions 
are point-like and trapped
in separate arms of the interferometer, i.e. $\psi_a(x)\psi_b(x)=0$ for all $x$. Using an initial $N$-particle two-mode state
$|\psi_{\mathrm{in}}\rangle=\sum_jc_j|j,N-j\rangle$, where $|c_j|^2$ is the probability of having $j$ atoms in mode $a$ and $(N-j)$ in $b$, we get
\begin{equation}
  \Delta^2\theta^{(m)}_{\rm ML}=\frac1m\frac{\Delta^2\hat J_z\cos^2\theta+\Delta^2\hat J_x\sin^2\theta}{\langle\hat J_x\rangle^2\cos^2\theta}.\label{sens_mzi}
\end{equation}
The details of the derivation are presented in \ref{ap_mzi}.
The above expression coincides with the phase uncertainty for the estimation of the phase $\theta$ from the average population imbalance between the two arms of the MZI. 
This coincidence can be explained as follows. When  atoms are trapped in two separate arms of the interferometer, their positions 
can be directly translated to the number of particles in each arm, without any loss of information. It is then not surprising that the estimators basing on 
the average population imbalance and the average density are equivalent.

Note that for a particular case of $\theta=0$, we get 
\begin{equation}
  \Delta^2\theta^{(m)}_{\rm ML}=\frac1{mN}\xi_n^2,\label{mzi_0}
\end{equation} where $\xi_n=\sqrt{N\frac{\Delta^2\hat J_z}{\langle\hat J_x\rangle^2}}$ is the spin-squeezing parameter \cite{wine,kita}
related to number squeezing of the initial state. The expression (\ref{mzi_0}) can provide up to the Heisenberg scaling of the phase uncertainty, once the interferometer is fed
with a strongly number-squeezed state \cite{pezze}.

\section{Estimation from the one-body density with the interference pattern}
\label{interf}

We now turn the attention to our case of interest, to which we devote the rest of the manuscript.
We consider the whole phase acquisition sequence as consiststing of two steps.
First is the phase-imprint performed in absence of two-body interactions, which in the two-mode picture is represented by the unitary operator $e^{-i\theta\hat J_z}$, and gives
\begin{equation}\label{impr}
  \hat\Psi(x|\theta)=\psi_a(x)e^{i\frac\theta2}\hat a+\psi_b(x)e^{-i\frac\theta2}\hat b.
\end{equation}
Then, the trap is opened. Since the two-body collisions are assumed to be not present, the two wave-packets freely spread and interfere. 
Atom interactions can strongly influence the expansion at an early stage, when the density of the clouds is high \cite{masiello,ceder}.
We assume that initially $\psi_{a/b}(x)$ have identical shape, but are centered around $\pm x_0$.
After long expansion time (in the so-called ``far field'') $\psi_{a/b}(x)\simeq e^{\mp i\frac{\kappa x}2}\cdot\tilde\psi\left(\frac x{\tilde\sigma^2}\right)$,
where $\tilde\sigma=\sqrt{\frac{\hbar t}{\mu}}$ ($\mu$ is the atomic mass and $t$ is the expansion time) and
$\kappa=2\frac{x_0}{\tilde\sigma^2}$, $\tilde\psi$ is the Fourier transform of the wave-packets at $t=0$, the same
for $\psi_{a/b}$ \cite{chwed,piazza}. Note that we have dropped the common factor $e^{i\frac{x^2}{2\tilde\sigma^2}}$.
Some aspect of this simple interferometric sequence, which does not require the implementation of a beam splitter, have been discussed in \cite{piazza}.

The field operator $\hat\Psi(x|\theta)$ gives $p_1(x_1|\theta)$ and $p_2(x_1,x_2|\theta)$ presented in the \ref{ap_pat}, which are put into Eq.~(\ref{sens}). 
The integrals are performed analytically assuming that the interference pattern consists of many fringes, giving
\begin{equation}\label{fin}
  \Delta^2\theta^{(m)}_{\rm ML}=\frac1{mN}\left[\xi_\phi^2+\frac{\sqrt{1-\nu^2}}{\nu^2}\right],
\end{equation}
where $\xi_\phi=\sqrt{N\frac{\Delta^2\hat J_y}{\langle\hat J_x\rangle^2}}$ is the spin-squeezing parameter \cite{grond} related to phase-squeezing of the initial state.
Also, we have introduced $\nu=\frac2N\langle\hat J_x\rangle$ -- i.e. the visibility of the interference fringes, see Eq.~(\ref{p1}) and below for details.
One important property of Eq.~(\ref{fin}) is its independece of the actual value of the phase $\theta$. This can be understood as follows. The probabilities 
(\ref{single}) and (\ref{double}) depend on $\theta$ in the same manner, i.e. via a sine or cosine function, see Eqs (\ref{p1}) and (\ref{p2}). 
  Since the integration in (\ref{c}) and (\ref{f1}) runs over the whole space, the shift of the trigonometric functions by a common factor $\theta$ is irrelevant, since
  within the envelope $\tilde\psi\left(\frac x{\tilde\sigma^2}\right)$ there are many interference fringes. 
  On the other hand, the sensitivity of the MZI (\ref{sens_mzi}) depends on $\theta$ and has some optimal ``working points'' because the wave-packets
  $\psi_{a/b}(x)$ do not add up to a wide envelope and therefore the argument for $\theta$-independence valid for the interference pattern does not apply here.
  Therefore, the $\theta$-independence of Eq.~(\ref{fin}) might be an advantage of this estimation protocol
  with respect to the momentum formula of the MZI, valid when the phase is estimated from the one-body density.

Note that the phase uncertainty of the MZI [Eq.~(\ref{mzi_0})] would closely resemble the above result if the second term of Eq.~(\ref{fin}) were absent. 
In the former case, $\Delta^2\theta^{(m)}_{\rm ML}$ benefits from the number-squeezing because the estimator is equivalent to the average population imbalance
between the arms of the interferometer, therefore a state with reduced population imbalance fluctuations decreases the uncertainty $\Delta^2\theta^{(m)}_{\rm ML}$.
Analogously, in the latter case, the decreasing fluctuations of the relative phase between the two modes would improve the estimation precision.

However, the situation complicates due to the presence of the second term in Eq.~(\ref{fin}). 
Namely, when $\xi_\phi$ drops, so does the fringe visibility $\nu$ and so the amount of information about the phase $\theta$ contained in the one-body density degrades. As
a consequence, the phase uncertainty of the estimation from $p_1(x|\theta)$ declines.

\begin{figure}[htb]
  \includegraphics[scale=0.45,clip]{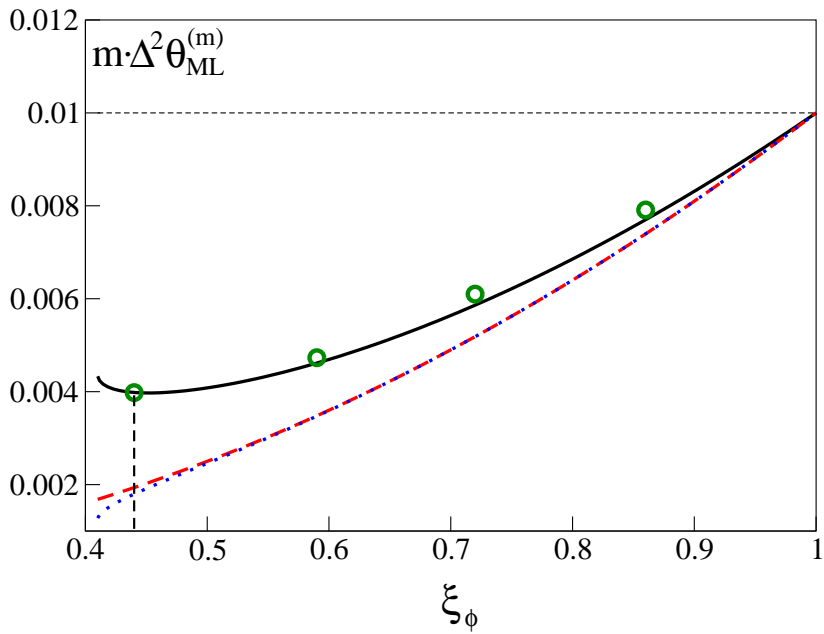}
  \caption{The phase uncertainty (\ref{fin}) -- black solid line -- as a function of $\xi_\phi$ for $N=100$ particles. The dashed red line is the phase uncertainty from the squeezing
    parameter (in absence of the second term in Eq.~(\ref{fin})), 
    while the dotted blue line is given by the inverse of the QFI. The horizontal dashed line denotes the SNL and the vertical dashed line indicates the position
    of the optimal point. The open green circles are the result of a numerical experiment (see text)
    for four values of $\xi_\phi=0.44,\ 0.59,\ 0.72$ and 0.86 with $m=10$, $n_{\rm rep}=4000$. 
  }
  \label{sq}
\end{figure}

It is now important to check whether, due to this interplay between improvement from the phase-squeezing and deterioration from the loss of visibility, Eq.~(\ref{fin}) 
can give SSN phase uncertainty at all. To this end, 
we calculate $\Delta^2\theta^{(m)}_{\rm ML}$ with phase-squeezed states for $N=100$, which  we generate by computing the ground state 
of the two-mode Hamiltonian $\hat H=-\hat J_x+\frac{U}{NJ}\hat J_z^2$ for
negative values of the interaction-to-tunelling ratio $\frac{U}{NJ}$. 
For a detailed study of the preparation of phase-squeezed states, including experimental sources of noise, see \cite{grond}.
For every value $\frac{U}{NJ}$, we find $|\psi_{\rm in}\rangle$ and calculate
$\xi_\phi$ and $\nu$. These values are substituted into Eq.~(\ref{fin}) and the resulting $\Delta^2\theta^{(m)}_{\rm ML}$ is plotted in Fig. \ref{sq} as a function of $\xi_\phi$.

Clearly, the uncertainty drops below the SNL. 
We notice the presence of an optimal point, where the gain from the spin-squeezing is balanced by the loss of visibility.
The figure also shows that the phase uncertainty (\ref{fin}) does not saturate the bound set by the Quantum Fisher Information (QFI), 
$\Delta^2\theta_{\rm QFI}\equiv\frac1m\frac1{F_Q}=\frac1m\frac1{4\Delta^2\hat J_z}$ \cite{nota_qfi}.

\begin{figure}[htb]
  \includegraphics[scale=0.45,clip]{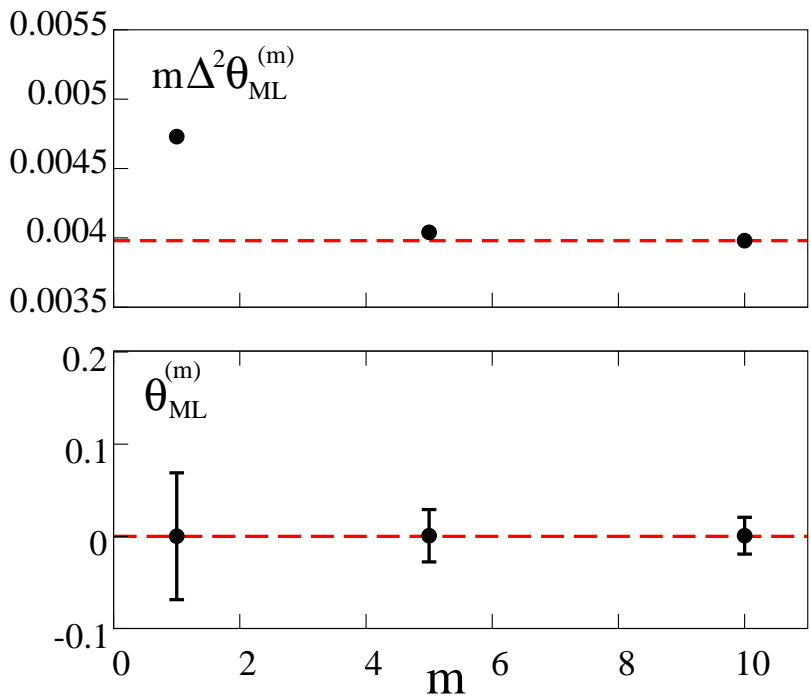}
  \caption{     
    The upper panel shows that the phase uncertainty of the ``experiment'' (black points) at the optimal point $\xi_\phi=0.44$ and $N=100$ beats the SNL (which is equal to 
    $m \Delta^2 \theta_{\rm ML}^{(m)}=0.01$ here, cf. Fig.~\ref{sq}) already for $m=1$ and 
    converges to the theoretical value (dashed red line) for $m=10$. The lower panel depicts the corresponding average plus error bars indicating uncertainties, while the true value of $\theta$ is indicated by the dashed red line.}
  \label{statistics}
\end{figure}

To support these analytical results for the asymptotics, we also study the full statistics of the protocol by simulating a phase-estimation experiment with $N=100$ particles. We generate the input state with 
a desired amount of phase-squeezing, evaluate the full $N$-body probability $p_N(\vec x_N|\theta)$, 
with which we draw a single realization yielding the $N$ positions. 
We repeat the experiment $m$ times and obtain one value of the phase $\theta^{(m)}_{\rm ML}$ using MLE with Eq.~(\ref{like}). This cycle is performed $n_{\rm rep}$ times and
the variance of the estimator is calculated on the resulting ensemble. The results, plotted (empty circles) in Fig. \ref{sq} 
for four values of $\xi_\phi$, $m=10$, and $n_{\rm rep}=4000$, agree with the theoretical value calculated with Eq.~(\ref{fin}).
An important information in view of an experimental implementation is that though formally the MLE saturates (\ref{fin}) when $m\to\infty$, in practice
$m=10$ is sufficient to reach the bound, as shown in the upper panel of Fig. \ref{statistics}. The lower panel of Fig.~\ref{statistics} shows instead the average value of the estimated phase plus uncertainties as a function of $m$. Here we have chosen the true value of $\theta$ to be zero, thus the figure show that the estimator is unbiased.

Our next step is to find the best scaling of Eq.~(\ref{fin}) with $N$. In order to get an analytical estimate, we model $|\psi_{\rm in}\rangle$ with a Gaussian, see \ref{ap_gaus}, and find that at the optimal point $m\Delta^2\theta_{\rm ML}^{(m),{\rm opt}}=2N^{-4/3}$. This prediction is compared with numerical results, where for every $N$ we evaluate the phase uncertainty (\ref{fin}) at the optimal state.
As shown in the inset of the Fig. \ref{err}, the agreement between the numerics and the 
Gaussian approximation is very good. Also, we numerically obtain the scaling of the QFI at the optimal state 
$m\Delta^2\theta_{\rm QFI}^{\rm opt}=N^{-4/3}$, which differs from $\Delta^2\theta_{\rm ML}^{(m),{\rm opt}}$ just by a factor of 2. 
Next, we discuss some characteristic experimental imperfections which can spoil the phase uncertainty (\ref{fin}).

\subsection{Impact of detection imperfections}
\label{noise}

\begin{figure}[htb]
  \includegraphics[scale=0.45,clip]{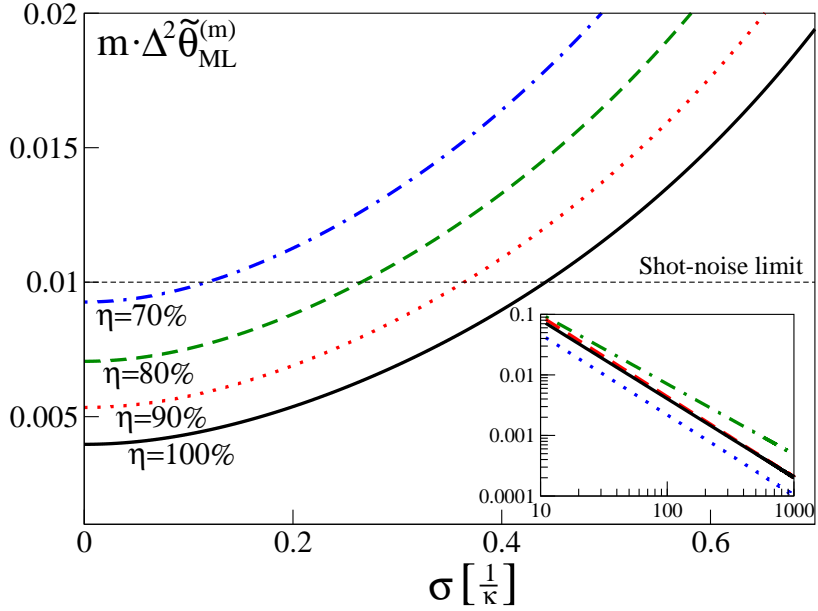}
  \caption{The phase uncertainty (\ref{sens_err}), calculated for $N=100$ in the optimal state (denoted by the vertical dashed line in Fig. \ref{sq}), 
    as a function of $\sigma$ for four different values of $\eta$. The horizontal dashed line denotes the SNL.
    The inset shows the phase uncertainty Eq.~(\ref{fin}) 
    as a function of $N$ (solid black line) with the optimal state, compared to the
    scaling from the Gaussian approximation, $m\Delta^2\theta_{\rm ML}^{(m),{\rm opt}}=2N^{-4/3}$ (dashed red line), the 
    scaling of the QFI $m\Delta^2\theta_{\rm QFI}^{(m),{\rm opt}}=N^{-4/3}$ (dotted blue line)
    and the scaling of Eq.~(\ref{sens_err}) with $\sigma=\frac{0.2}\kappa$ and $\eta=90\%$, giving $m\Delta^2\tilde\theta_{\rm ML}^{(m),{\rm opt}}=1.48N^{-1.16}$ (dash-dotted green line).}
  \label{err}
\end{figure}

Single-particle detection, which is the basis of the estimation scheme discussed in Sec.~\ref{sing_at_det}, 
is affected by two dominant sources of noise: limited efficiency and finite spatial resolution, see Fig.~\ref{scheme}. 
The former is incorporated by letting the index $k$ run from 1 to $n<N$ in Eq.~(\ref{like}). 
In effect, $n$ replaces $N$ in Eq.~(\ref{sens}). The finite spatial resolution modifies instead both $C$ and $F_1$. We implement it by convoluting the probabilities $p_1(x_1|\theta)$ 
and $p_2(x_1,x_2|\theta)$ with $\tilde p(x_1|x_1')=\frac1{\sqrt{2\pi\sigma^2}}e^{-(x_1-x_1')^2/(2\sigma^2)}$ (and analogously for $x_2$) -- a Gaussian
probability of detecting an atom at $x_1$ given its true position $x_1'$. The convolutions are calculated analytically and the phase uncertainty becomes
\begin{equation}\label{sens_err}
  \Delta^2\tilde\theta^{(m)}_{\rm ML}=\frac1{mN}\left[\xi_\phi^2+\frac{\left(\sqrt{1-\nu^2e^{-\kappa^2\sigma^2}}+1\right)e^{\kappa^2\sigma^2}-\eta}{\eta\nu^2}\right],
\end{equation}
where $\kappa=2\frac{x_0}{\tilde\sigma^2}$ and $\tilde\sigma=\sqrt{\frac{\hbar t}{\mu}}$ were defined below Eq.~(\ref{impr}). 
Above, we assumed that $N,n\gg1$, $\eta=\frac nN$ and the tilde denotes the phase uncertainty in presence of errors. Note that for $\eta=100\%$ and $\sigma=0$ we recover Eq.~(\ref{fin}). 
In Fig. \ref{err} we plot Eq.~(\ref{sens_err}) as a function of $\sigma$
for $N=100$ and $\eta=100\%,90\%,80\%$ and 70\%. 
For instance, with resolution of 1/30-th of a fringe ($\sigma=0.033\times\frac{2\pi}\kappa\simeq\frac{0.2}\kappa$) and $\eta=90\%$, 
the SSN scaling is $m\Delta^2\tilde\theta^{(m),{\rm opt}}_{\rm ML}=1.48N^{-1.16}$, shown in the inset.
Also note that even in presence of the noise, the phase uncertainty (\ref{sens_err}) does not depend on $\theta$.

The two main sources of detection noise affecting the least-squares fit estimation protocol defined in Sec.~\ref{mult_at_det}, 
not shown in Fig. \ref{scheme}, are  imperfect atom counting and the finite bin size.
To model the former, we assume that the number of atoms in each bin is measured with some uncertainty, and convolute the probabilities entering the 
fit likelihood function with a Gaussian error distribution, $p_{\rm err}(\bar{n}_k|\bar{n}_k')=\frac{\sqrt m}{\sqrt{2\pi\sigma_{\rm err}^2}}e^{-(\bar{n}_k-\bar{n}_k')^2/(2\sigma_{\rm err}^2/m)}$. 
As a result, the variance is increased by $\sigma_{\rm err}^2/m$. 
Very promising detection techniques 
based on the detection of fluorescence photons rely upon detection of on average  $\alpha$ fluorescence photons per atom,
and this number fluctuates at the shot noise level. Error propagation gives $\sigma_{\rm err}^2=\frac1\alpha\langle n_k\rangle$ and the phase uncertainty (\ref{sens_fit}) 
becomes
\begin{equation*}
  \Delta^2\tilde{\theta}^{(m)}_{\rm fit}=\Delta^2{\theta}^{(m)}_{\rm fit}
  +\frac1m\frac{1}{\alpha}\frac1{\sum_{k=1}^{n_{\rm bin}} \frac{\left[\partial_\theta\langle n_k\rangle\right]^2}{\langle n_k \rangle}}.
\end{equation*}
Using $N=100$ and the optimal state denoted by the vertical line in Fig. \ref{sq}, 
for the bin size $\Delta x=\frac{0.2}\kappa$ a SSN phase uncertainty is preserved for $\alpha\gtrsim2.2$ -- a condition
well satisfied by the light-sheet technique, which can give $\alpha\simeq10$  photons per atom \cite{perrin_sheet}.

\section{Conclusions}
\label{conc}



We have derived an expression for the phase estimation uncertainty for a generic situation 
where the phase is inferred from the positions of probe particles, when only the one-body density is known.
In a Mach-Zehnder type interferometer, the sensitivity of this protocol coincides with the well known error propagation formula and 
we recover the known Heisenberg limited phase uncertainty, $\Delta\theta\propto N^{-1}$. Then 
 we consider the simplest ``double-slit'' interferometer based on spatially interfering wave-packets suggested in Ref. \cite{piazza}, which still scales at best as
$\Delta\theta\propto N^{-2/3}$, limited by the loss of fringe visibility.
Nevertheless, the phase uncertainty for the interference pattern (\ref{fin}) has a major advantage over the MZI (\ref{sens_mzi}). Namely,
it performs equally well for any value of $\theta$, while (\ref{sens_mzi}) can reach very high values around $\theta=\frac\pi2$ and $\theta=\frac32\pi$.

The interferometric protocol employing the interference pattern could be implemented with a BEC trapped in a double-well potential.
After imprinting the phase and switching the trap off, the two clouds would expand and interfere. The atoms could then be detected using for instance the light-sheet method 
\cite{perrin_sheet}, based on the fluorescence measurement of photons scattered by the atoms crossing a laser beam. Another possible scheme relies upon letting
the atoms fall onto an optical lattice. If the interference pattern is dilute so that there is no more than one atom per site, 
their positions could be detected by a fluorescence measurement \cite{sherson} with ultra-high efficiency and resolution.
Recent quantum interferometry experiments \cite{gross, riedel, klempt} indicate that very important effects limitating the phase uncertainty are the detection imperfections, 
which we believe to have realistically taken into account in our proposal. Another relevant constraint to the precision of a double-well interferometer comes from the
noise present in the interferometric sequence i).
A recent theoretical work \cite{grond} shows that the amount of squeezing at the optimal point
($\xi_\phi=0.44$ for $N=100$ particles) could be reached with a double-well BEC using a refocusing method  even in presence of the latter source of noise.


\section{Acknowledgements}
We gratefully acknowledge discussions with Julian Grond and Aur\'elien Perrin. 
We thank Baltazar Brukalski for preparing Fig.\,\ref{scheme}. J. Ch. acknowledges Foundation for Polish Science International TEAM Program cofinanced 
by the EU European Regional Development Fund and the support of the National Science Center. 
P.H. acknowledges support from the ERC Starting Grant GEDENTQOPT and CHIST-ERA QUASAR.
F.P. acknowledges support by the Alexander Von Humboldt foundation. A.S. acknowledges support of the EU-STREP Project QIBEC.
\appendix

\section{Derivation of the phase uncertainty}
\label{ap_lik}
\subsection{Single-atom detection}
The proof that the estimator taken as the phase $\theta_{\rm ML}^{(m)}$ 
which maximizes the likelihood function of Eq.~(\ref{like}) is unbiased and 
has the variance of Eq.~(\ref{sens}) asymptotically in $m$ can be performed
along the lines of the standard proof 
that the Maximum-Likelihood method
saturates the Cram{\'e}r-Rao bound~\cite{Cramer_book} for $m\gg 1$
\begin{equation}\label{eq:CR}
  \Delta^2\theta_{\rm ML}^{(m)}\ge \frac{1}{mF},
\end{equation}
where $F$ is the Fisher information to be defined below.  
The original proof was given in Ref.~\cite{Cramer_book}. 
Recent formulation can be found in Ref.~\cite{Ferguson_book}, and a more accessible albeit less rigorous 
version is given in Ref.~\cite{Barlow_book}.

The proof consists of two steps. Firstly, we show that the estimator is {\em consistent}, {\em i.e.},
for $m\to\infty$, the probability that $\theta_{\rm ML}^{(m)}\neq\theta$
goes to zero, which means
that the estimator approaches the true value of the phase shift
asymptotically. This demonstration follows Ref.~\cite{Ferguson_book}. Secondly, we adapt the simplified
proof of Ref.~\cite{Barlow_book} (see also \cite{Cramer_book}) 
that a consistent ML estimator is also {\em efficient}, which means that it 
saturates the Cram{\'e}r-Rao bound, Eq.~(\ref{eq:CR}), for $m\gg 1$.
In particular, the proof shows that for large $m$, $\theta_{\rm ML}^{(m)}$ 
is distributed with a Gaussian distribution with 
the variance of Eq.~(2) and mean $\theta$, hence it is also unbiased. 

Before we start, a remark is in order. 
It might seem surprising or even wrong
to use the likelihood function of Eq.~(\ref{like}). The reason is that we use only the 
single particle probabilities to estimate the phase, even though in
a single shot of the experiment, the $N$ particles will be correlated
in general. Hence the probability that in a single shot the second particle
arrives at a position $x_2$ generally depends on the position $x_1$ where
the first particle was detected. In traditional ML estimation, one would 
therefore define the likelihood function as 
\begin{equation}
  {\cal L}(\varphi)=\prod_{i=1}^m p_N(\vec x^{(i)}_N|\varphi)
\end{equation}
with the $N$-particle conditional probability density $p_N(\vec x^{(i)}_N|\varphi)$
instead and define the estimator as the maximum of this function. 
As mentioned above it can be shown~\cite{Cramer_book,Ferguson_book,Barlow_book}
that this estimator is consistent, unbiased, and that it saturates the Cramer-Rao 
bound with the Fisher information
\begin{equation}\label{eq:F_N}
  F_N=\int d{\vec x} \frac{1}{p_N(\vec x_N|\theta)}\Big(\partial_\theta p_N(\vec x_N|\theta)\Big)^2. 
\end{equation}
This bound cannot be overcome by any other estimator using the
results of the measurement governed by the probability density 
$p_N(\vec x_N|\theta)$ or by any of its reductions. 

However, there are no restrictions on how an estimator can be defined,
there will only be differences in the performance of different
estimators. We define the estimator based on the single-particle probability 
density only. As the proof below shows, this
estimator is consistent, unbiased, and has the variance of Eq.~(2).
The variance is ultimately limited (but is generally larger than)
the ultimate limit from Eq.~(\ref{eq:CR}) with the Fisher information from 
Eq.~(\ref{eq:F_N}). However, it has the advantage that it is 
accessible experimentally and also allows for sub shot-noise phase estimation,
as shown in the main article.

{\em Consistency}.
We recall the definition of $\theta_{\rm ML}^{(m)}$,
which is the value of the parameter $\varphi$ which maximizes 
${\cal L}(\varphi)$ from Eq.~(1) from the main article.
Equivalently, it maximizes
\begin{equation}\label{eq:single}
  f^{(m)}(\varphi) \equiv\frac{1}{m} \sum_{i=1}^m  \sum_{k=1}^N \ln p_1(x_k^{(i)}|\varphi),   
\end{equation}
where $N$ is the number of particles and $m$ the number of independent repetitions of the experiment. 
The events $\vec x_N$ are distributed with the conditional probability density $p_N(\vec x_N|\theta)$, where $\theta$ is the true value of the phase shift, and 
$p_1(x_k|\theta)$ is obtained from $p_N(\vec x_N|\theta)$ by integrating over all $x_{j\neq k}$.

We assume {\em identifiability}, {\em i.e.}, that $p_1(x|\theta)=p_1(x|\theta')$ for all $x$ is equivalent
to $\theta=\theta'$. Consistency is then proved by showing that $f(\varphi) = \lim \limits_{m \to \infty} f^{(m)}(\varphi)$ has a maximum at $\varphi = \theta$ as follows:
\begin{eqnarray}
  &&f(\varphi) - f(\theta)=\lim \limits_{m \to \infty} \frac{1}{m} \sum_{i=1}^m  \sum_{k=1}^N
  \Big(\ln p_1(x_k^{(i)}|\varphi) - \ln p_1(x_k^{(i)}|\theta) \Big)  \nonumber\\
  &&=\sum_{k=1}^N \int d\vec x_N\, p_1(x_k|\theta)\Big( \ln p_1(x_k|\varphi) - \ln p_1(x_k|\theta) \Big)   \nonumber \\ 
  &&=N \int dx\,p_1(x|\theta) \ln \frac{p_1(x|\varphi)}{p_1(x|\theta)} 
  \le 0, \label{eq:consistency}
\end{eqnarray}
where we have used $\ln(y) \le y -1$. The equality is obtained {\it iff} $y=1$.
Hence the inequality Eq.~(\ref{eq:consistency}) is saturated {\em iff}
$p_1(x|\varphi)=p_1(x|\theta)$ for all $x$. It follows that $\varphi = \theta$
by the identifiability assumption. Hence $\theta^{(m)}_{\rm ML} \to \theta$ for $m\to \infty$.

{\em Efficiency}.
We expand the first derivative of ${\cal L}$ from Eq.~(1) of the main
article around $\theta$, 
\begin{equation} \partial_\varphi\log\mathcal{L}(\varphi)\simeq\partial_\varphi\log\mathcal{L}(\varphi)\big\vert_{\theta}+\partial^2_\varphi\log\mathcal{L}(\varphi)\big\vert_{\theta}(\varphi-\theta).
\end{equation}
We now set $\varphi=\theta_{\rm ML}^{(m)}$. Since this phase maximizes ${\cal L}$, 
the left-hand-side vanishes and we obtain
\begin{equation}\label{approx} (\theta_{\rm ML}^{(m)}-\theta)\simeq-\frac{\partial_\varphi\log\mathcal{L}(\varphi)\big\vert_{\theta}}{\partial^2_\varphi\log\mathcal{L}(\varphi)\big\vert_{\theta}}.
\end{equation}
The consistency of the estimator ensures that we can neglect
terms of higher order in $\theta_{\rm ML}^{(m)}-\theta$ provided that 
$m$ is large enough.

In order to investigate how $\theta_{\rm ML}^{(m)}-\theta$ is distributed,
we start by computing the average of the denominator,
\begin{eqnarray}\label{denom}
&&\partial^2_\varphi\log\mathcal{L}(\varphi)\big\vert_{\theta}=m\left[\frac1m\sum_{i=1}^m\sum_{k=1}^N\partial_\varphi^2\log \big(p_1(x^{(i)}_k|\varphi)\big)\Big\vert_{\theta}\right]
  \nonumber\\
  &&\rightarrow_{m\gg1}m \sum_{k=1}^N\int d\vec x_N p_N(\vec x_N|\theta)\partial_\varphi^2\log \big(p_1(x_k|\varphi)\big)\Big\vert_{\theta}\nonumber\\
  &&=-mNF_1.
\end{eqnarray}
The coefficient $N$ results from the indistinguishability of the particles and $F_1$ is the Fisher information calculated with the single particle probability density,
\begin{equation}
  F_1=\int dx_1\frac1{p(x_1|\theta)}\big(\partial_\varphi p(x_1|\varphi) |_{\theta}\big)^2.
\end{equation}
Coming back to Eq.~(\ref{approx}), we get
\begin{equation} (\theta_{\rm ML}^{(m)}-\theta)\simeq\frac1m\sum_{i=1}^m\left[\frac1{NF_1}\sum_{k=1}^N\partial_\varphi \log \big(p_1(x^{(i)}_k|\varphi)\big)\Big\vert_{\theta}\right].
\end{equation}
Hence the difference $\theta_{\rm ML}^{(m)}-\theta$ is the average of $m$ random variables, which in the central limit are distributed with a Gaussian probability. 
With a calculation similar to the one of Eq.~(\ref{denom}) one obtains that the average value vanishes, which means that the MLE is \emph{unbiased}. 
The variance of the distribution in the central limit is 
\begin{equation} 
  \Delta^2\theta_{\rm ML}^{(m)}=\frac{1}{m}\frac1{NF_1}\left(1+(N-1)\frac C{F_1}\right)
\end{equation}
where 
\begin{equation} 
  C=\int dx_1 dx_2\,p_2(x_1,x_2|\theta)\frac{\partial_\theta p_1(x_1|\theta)}{p_1(x_1|\theta)}\frac{\partial_\theta p_1(x_2|\theta)}{p_1(x_2|\theta)}.
\end{equation}
Therefore the Fisher information from Eq.~(\ref{eq:CR}) is
\begin{equation}
  F=\frac{NF_1}{\left(1+(N-1)\frac C{F_1}\right)}.
\end{equation}
Hence the correlations enter {\em via} the two-particle correlation function $p_2$ even though in the definition of the estimator 
only the single body density $p_1$ is used.

\subsection{Multiple-atom detection}

The multiple-atom detection relies upon dividing the space into $n_{\rm bin}$ bins, each of size $\Delta x$. 
In every bin, the number of atoms is measured and this result is averaged over $m$ realizations, giving
\begin{equation}
  \bar n_k=\frac{1}{m}\sum_{i=1}^mn_k^{(i)}.
\end{equation}
According to the central limit theorem, for large $m$ the 
probability of detecting $\bar n_k$ is Gaussian, 
\begin{equation}
  p(\bar n_k|\varphi)=\frac1{\sqrt{2\pi\Delta^2n_k/m}}e^{-\frac{(\bar n_k-\langle n_k\rangle)^2}{2\Delta^2n_k/m}}.
\end{equation}
Here, $\langle n_k\rangle=\lim\limits_{m\rightarrow\infty}\bar n_k$ and $\Delta^2n_k$ are the associated fluctuations and both depend on the value of $\varphi$.
We construct the likelihood function as follows
\begin{equation}
  \mathcal{L}_{\rm fit}(\varphi)=\prod_{k=1}^{n_{\rm bin}}p(\bar n_k|\varphi).
\end{equation}
Again, we follow the steps of the proof of Fisher's theorem and expand the derivative of the logarithm of the likelihood function 
around the true value and set $\varphi=\theta^{(m)}_{\rm ML}$, 
\begin{equation}
  (\theta^{(m)}_{\rm ML}-\theta)\simeq-\frac{\partial_\varphi\log\mathcal{L}_{\rm fit}(\varphi)\vert_{\theta}}{\partial^2_\varphi\log\mathcal{L}_{\rm fit}(\varphi)\vert_\theta}.\label{diff}
\end{equation}
For the following calculations we introduce the more compact notation 
\begin{equation}\label{def}
  \partial_\varphi u(\varphi)\vert_{\theta}\equiv\partial_{\theta} u,
\end{equation}
used also in the main text. 
We start with calculating the denominator of Eq.~(\ref{diff}) which in the large $m$ limit reads,
  \begin{eqnarray}
    \partial^2_\theta\log\mathcal{L}_{\rm fit}=-m\sum_{k=1}^{n_{\rm bin}}\left[\frac{(\partial_\theta\langle n_k\rangle)^2}{\Delta^2 n_k}+\left[\langle n_k\rangle-\bar n_k\right]
      \left(\frac{\partial^2_\theta\langle n_k\rangle}{\Delta^2 n_k}+\partial_\theta\langle n_k\rangle\partial_\theta\frac1{\Delta^2n_k}\right)\right.\nonumber\\
      +\left.\frac12\left[\bar n_k-\langle n_k\rangle\right]^2\partial^2_\theta\frac1{\Delta^2n_k}\right].
  \end{eqnarray}
According to (\ref{def}), both $\langle n_k\rangle$ and $\Delta^2n_k$ in the above equation are a function of $\theta$. 
In analogy to Eq.~(\ref{denom}) for $m\rightarrow\infty$ above denominator is replaced with its average value.
Upon averaging, the second term vanishes and the third term is proportional to $\Delta^2n_k/m$, which is negligible in the limit of large $m$. Therefore we obtain
\begin{equation}
  \partial^2_\theta\log\mathcal{L}_{\rm fit}\simeq -m\sum_{k=1}^{n_{\rm bin}}\frac{(\partial_\theta\langle n_k\rangle)^2}{\Delta^2n_k}.
\end{equation}
The average of the square of the nominator of Eq.~(\ref{diff}) reads
  \begin{eqnarray}
    \nonumber\big\langle(\partial_\theta\log\mathcal{L}_{\rm fit})^2\big\rangle\simeq m^2\sum_{k,l=1}^{n_{\rm bin}}
    \frac{\partial_\theta \langle n_k\rangle\partial_\theta\langle n_l\rangle}{\Delta^2n_k\Delta^2n_l}\Big\langle(\bar n_k-\langle n_k\rangle)(\bar n_l-\langle n_l\rangle)\Big\rangle\\
    -2m^2\sum_{k,l=1}^{n_{\rm bin}}
    \frac{\partial_\theta \langle n_k\rangle}{\Delta^2n_k}\partial_\theta\frac1{2\Delta^2n_l}\Big\langle(\bar n_k-\langle n_k\rangle)(\bar n_l-\langle n_l\rangle)^2\Big\rangle\nonumber\\
    +m^2\sum_{k,l=1}^{n_{\rm bin}}
    \partial_\theta\frac1{2\Delta^2n_k}\partial_\theta\frac1{2\Delta^2n_l}\Big\langle(\bar n_k-\langle n_k\rangle)^2(\bar n_l-\langle n_l\rangle)^2\Big\rangle.
  \end{eqnarray}
The second term vanishes and the last, after the average is calculated in the central limit, becomes $m$-independent, and thus can be dropped when compared to the first term. 

Let us now separate the sum over $k,l$ into $k=l$ and $k\neq l$ parts. The first one simply gives
\begin{equation}
  m\sum_{k=1}^{n_{\rm bin}}\frac{(\partial_\theta\langle n_k\rangle)^2}{\Delta^2n_k}.
\end{equation}
The non-diagonal part $k\neq l$ depends on the two-site correlations $\sigma_{kl}^2=m\Big\langle(\bar n_k-\langle n_k\rangle)(\bar n_l-\langle n_l\rangle)\Big\rangle$. 
To justify Eq.~(5) of the main text, we notice that the atom-number fluctuations read
\begin{equation}
  \Delta^2n_k=p_1(x_k|\theta)\Delta x+(p_2(x_k,x_k|\theta)-p_1(x_k|\theta)^2)(\Delta x)^2
\end{equation}
and thus in a small bin-size limit are Poissonian, i.e.  $\Delta^2n_k=p_1(x_k|\theta)\Delta x=\langle n_k\rangle$. Therefore, 
\begin{equation}
  \Delta^2\theta^{(m)}_{\rm fit}=\frac{\sum_{k=1}^{n_{\rm bin}} \frac{\left[\partial_\theta\langle n_k\rangle\right]^2}{\langle n_k \rangle}+
    \sum_{k\neq l=1}^{n_{\rm bin}}\sigma_{k,l}^2\frac{\partial_\theta\langle n_k\rangle}{\langle n_k \rangle}\frac{\partial_\theta\langle n_l\rangle}{\langle n_l \rangle}}
    {m\left[\sum_{k=1}^{n_{\rm bin}} \frac{\left[\partial_\theta\langle n_k\rangle\right]^2}{\langle n_k \rangle}\right]^2}.\label{sens_fita}
\end{equation}
The last step is to calculate the cross-corelation term
\begin{eqnarray*}
  \fl\sigma_{k,l}^2&=&(\Delta x)^2\left[\langle\hat\Psi^\dagger(x_k|\theta)\hat\Psi(x_k|\theta)\hat\Psi^\dagger(x_l|\theta)\hat\Psi(x_l|\theta)\rangle
  -\langle\hat\Psi^\dagger(x_k|\theta)\hat\Psi(x_k|\theta)\rangle\langle\hat\Psi^\dagger(x_l|\theta)\hat\Psi(x_l|\theta)\rangle\right]\\
  \fl&=&(\Delta x)^2N\left[(N-1)p_2(x_k,x_l|\theta)-Np_1(x_k|\theta)p_1(x_k|\theta)\right],
\end{eqnarray*}
where in the last line we used
\begin{eqnarray}
  \langle\hat\Psi^\dagger(x_k|\theta)\hat\Psi(x_k|\theta)\hat\Psi^\dagger(x_l|\theta)\hat\Psi(x_l|\theta)\rangle=\nonumber\\
  \langle\hat\Psi^\dagger(x_k|\theta)\hat\Psi^\dagger(x_l|\theta)\hat\Psi(x_l|\theta)\hat\Psi(x_k|\theta)\rangle=N(N-1)p_2(x_k,x_l|\theta),
\end{eqnarray}
which is true for $k\neq l$.

\section{Derivation of the phase uncertainty for the MZI}
\label{ap_mzi}

When the two wave-packets are fully separated, so that $\psi_a(x)\psi_b(x)=0$ for all $x$, the one-body probability reads
\begin{displaymath}
p_1(x|\theta)=\frac1N\langle\hat\Psi^\dagger(x|\theta)\hat\Psi(x|\theta)\rangle=\frac 12\left[|\psi_a(x)|^2(1-\nu\sin\theta)+|\psi_b(x)|^2(1+\nu\sin\theta)\right],
\end{displaymath}
where $\nu=\frac2N\langle\hat J_x\rangle$ is the fringe visibility. This probability gives
\begin{eqnarray*}
  \fl F_1=\int dx\frac{\left[\partial_\theta p_1(x|\theta)\right]^2}{p_1(x|\theta)}
  =\frac{\nu^2\cos^2\!\theta}2
  \int dx\ \frac{|\psi_a(x)|^4+|\psi_b(x)|^4}{|\psi_a(x)|^2(1-\nu\sin\theta)+|\psi_b(x)|^2(1+\nu\sin\theta)}\\
  =\frac{\nu^2\cos^2\!\theta}{1-\nu^2\sin^2\!\theta}.
\end{eqnarray*}
Now we calculate the second order probability and get [cf. Eq.~(\ref{double})]
\begin{eqnarray*}
  \fl N(N-1)p_2(x_1,x_2|\theta)=
  \big\langle\hat\Psi^\dagger(x_1|\theta)\Psi^\dagger(x_2|\theta)\hat\Psi(x_2|\theta)\hat\Psi(x_1|\theta)\big\rangle=
  \frac{N^2}2(|\psi_a(x_1)|^2+|\psi_b(x_1)|^2)p_1(x_2)\\
  \fl +(|\psi_a(x_1)|^2-|\psi_b(x_1)|^2)(|\psi_a(x_2)|^2-|\psi_b(x_2)|^2)
  \left[\cos^2\theta\left(\langle\hat J_z^2\rangle-\frac N4\right)+\sin^2\theta\left(\langle\hat J_x^2\rangle-\frac N4\right)\right]\\
  \fl  +(|\psi_b(x_1)|^2-|\psi_a(x_1)|^2)(|\psi_a(x_2)|^2+|\psi_b(x_2)|^2)\frac N2\sin\theta\langle\hat J_x\rangle.
\end{eqnarray*}
We insert this function together with $p_1(x|\theta)$ into the definition of $C$ and obtain
\begin{eqnarray*} 
  N(N-1)\cdot C&=&\frac{N^2}4(\alpha+\beta)^2
  +\left(N-\frac12\right)\langle\hat J_x\rangle(\beta^2-\alpha^2)\sin\theta\\
  &+&
  \left(\langle\hat J_z^2\rangle\cos^2\theta+\langle\hat J_x^2\rangle\sin^2\theta-\frac N4\right)(\alpha-\beta)^2,
\end{eqnarray*}
where
\begin{equation}
  \alpha=-\frac{\langle\hat J_x\rangle\cos\theta}{\frac N2-\langle\hat J_x\rangle\sin\theta}\ \ \ \mathrm{and}\ \ \ 
  \beta=\frac{\langle\hat J_x\rangle\cos\theta}{\frac N2+\langle\hat J_x\rangle\sin\theta}.
\end{equation}
If we now combine expressions for $F_1$ and $C$ as in Eq.~(\ref{sens}), we obtain Eq.~(\ref{sens_mzi}).

\section{The one- and two-body probabilites for the interference pattern}
\label{ap_pat}

The explicit expression for the one-body probability comes directly from the definitons of $p_1(x|\theta)$ and the angular momentum operators,
\begin{equation}
  \fl p_1(x|\theta)=
  \frac1N\big\langle\hat\Psi^\dagger(x|\theta)\hat\Psi(x|\theta)\big\rangle=\Big\vert\tilde\psi\left(\frac{x}{\tilde\sigma^2}\right)
  \Big\vert^2\left(1+\frac{2\langle\hat J_x\rangle}N\cos(\kappa x+\theta)\right)\label{p1}.
\end{equation}
Clearly, the coefficent $\nu\equiv\frac{2\langle\hat J_x\rangle}N$ is the visibility of the interference fringes.
The two body probability reads [cf. Eq.~(\ref{double})]
\begin{eqnarray}\label{p2}
  \fl&&p_2(x_1,x_2|\theta)=\frac{\big\langle\hat\Psi^\dagger(x_1|\theta)\Psi^\dagger(x_2|\theta)\hat\Psi(x_2|\theta)\hat\Psi(x_1|\theta)\big\rangle}{N(N-1)}=
  \Big\vert\tilde\psi\left(\frac{x_1}{\tilde\sigma^2}\right)\Big\vert^2\Big\vert\tilde\psi\left(\frac{x_2}{\tilde\sigma^2}\right)\Big\vert^2\times\nonumber\\
  \fl&&\times
  \left[1- \frac1{(N-1)}\cos(\kappa(x_1-x_2))+\frac{2\langle\hat J_x\rangle}N[\cos(\kappa x_1+\theta)+\cos(\kappa x_2+\theta)]\right.\nonumber\\
  \fl&&\left.+\frac{4\langle\hat J_x^2\rangle}{N(N-1)}\cos(\kappa x_1+\theta)\cos(\kappa x_2+\theta)
  +\frac{4\langle\hat J_y^2\rangle}{N(N-1)}\sin(\kappa x_1+\theta)\sin(\kappa x_2+\theta)\right]\label{p2}.
\end{eqnarray}

\section{Gaussian scaling}
\label{ap_gaus}
To find the best possible scaling of the phase uncertainty (\ref{fin}) with $N$,
we model $|\psi_{\rm in}\rangle$ with a Gaussian state as follows
\begin{equation}
  |\psi_{\rm in}\rangle\propto e^{-i\frac\pi2\hat J_x}\sum_je^{-\frac{\left(j-\frac N2\right)^2}{N\cdot \xi_\phi}}|j,N-j\rangle.
\end{equation}
The operator $e^{-i\frac\pi2\hat J_x}$ represents a beam-splitter which transforms the number- to the phase-squeezed state.
For $\xi_\phi=1$ the resulting state is spin-coherent, and by decreasing $\xi_\phi$ we increase the amount of phase-squeezing. 

This state is used to calculate the expectation values $\langle\hat J_x\rangle$ and $\langle\hat J_y^2\rangle$ from Eq.~(\ref{fin}). 
For $N\gg1$ the summation over $j$ can be approximated with an integral. This way we obtain analytical expressions
\begin{equation}
 \langle\hat J_x\rangle^2=\frac{N^2}4e^{-\frac1{N\cdot\xi_\phi}}\ \ \ {\rm and}\ \ \ \langle\hat J_y^2\rangle=\frac{N^2}8\left(1-e^{-\frac2{N\cdot\xi_\phi}}\right).
\end{equation} 
which are then substituted into Eq.~(\ref{fin}). Taking $\xi_\phi=N^{-\beta}$, where $0\leq\beta\leq1$, we get 
\begin{equation}
  m\Delta^2\theta^{(m)}_{\rm ML}=N^{-(\beta+1)}+N^{\frac{\beta-3}2}. 
\end{equation}
The phase uncertainty is optimal, when these two terms are equal, otherwise one of them would dominate at large $N$. This condition gives $\beta^{\rm opt}=\frac13$ 
and $m\Delta^2\theta_{\rm ML}^{(m),{\rm opt}}=2N^{-4/3}$.


\begin{thebibliography}{40}
\bibitem{nota_der}{We use an abbreviated notation for the derivatives, $\partial_\theta f=\partial_\theta f(\theta)\equiv\left[\partial_\varphi f(\varphi)\right]\big|_{\varphi=\theta}$.}
\bibitem{nota_J}
  {
    The  angular momentum operators are
    $\hat J_x \equiv (\hat a^\dagger\hat b^{\phantom{\dagger}}+
    \hat b^\dagger\hat a^{\phantom{\dagger}})/2$, $\hat J_y \equiv (\hat a^\dagger\hat b^{\phantom{\dagger}}-
    \hat b^\dagger\hat a^{\phantom{\dagger}})/2i$ and $\hat J_z \equiv (\hat a^\dagger\hat a^{\phantom{\dagger}}-
    \hat b^\dagger\hat b^{\phantom{\dagger}})/2$.
  }
\end{thebibliography}
\end{document}